\documentclass[12pt,nofootinbib,aps,amsmath,amssymb,preprint,showpacs]{revtex4}

\usepackage[T1]{fontenc} \usepackage[latin1]{inputenc}
\usepackage{amsmath,color,ulem} \usepackage{graphicx}
\usepackage{amssymb}
\newcommand\underrel[2]{\mathrel{\mathop{#2}\limits_{#1}}}
\usepackage{tabulary}
\newcolumntype{K}[1]{>{\centering\arraybackslash}p{#1}}
\begin{document}

\title{Generalised Langevin Equation Formulation for Anomalous
  Diffusion in the Ising Model at the Critical Temperature}

\author{Wei Zhong$^\dagger$} 
\email{w.zhong1@uu.nl}
\author{Debabrata Panja$^\dagger$}
 \author{Gerard T. Barkema$^\dagger$}
 \author{Robin C. Ball$^\ddagger$}
\affiliation{
 $^\dagger$Department of Information and Computing Sciences, Utrecht University, Princetonplein 5, 3584 CC Utrecht, The Netherlands\\
 $^\ddagger$Department of Physics, University of Warwick, Coventry CV4 7AL, United Kingdom
}

\date{\today} 

\begin{abstract}
We consider the two- (2D) and three-dimensional (3D) Ising model on a
square lattice at the critical temperature $T_c$, under Monte-Carlo spin flip
dynamics. The bulk magnetisation and the magnetisation of a tagged
line in the 2D Ising model, and the bulk magnetisation and the
magnetisation of a tagged plane in the 3D Ising model exhibit anomalous
diffusion. Specifically, their mean-square displacement increases as
power-laws in time, collectively denoted as $\sim t^c$, where $c$ is
the anomalous exponent. We argue that the anomalous diffusion in all
these quantities for the Ising model stems from time-dependent
restoring forces, decaying as power-laws in time --- also with
exponent $c$ --- in striking similarity to anomalous diffusion in
polymeric systems. Prompted by our previous work that has established
a memory-kernel based Generalised Langevin Equation (GLE) formulation
for polymeric systems, we show that a closely analogous GLE
formulation holds for the Ising model as well. We obtain the memory
kernels from spin-spin correlation functions, and the formulation
allows us to consistently explain anomalous diffusion as well as
anomalous response of the Ising model to an externally applied
magnetic field in a consistent manner.
\end{abstract}

\maketitle

\section{Introduction \label{sce1}}

In case of normal diffusion the mean-square displacement (msd) of a
particle $\langle\Delta r^2(t)\rangle$ increases linearly in time. The
term anomalous diffusion is used to denote a particle's mean-square
displacement $\langle\Delta r^2(t)\rangle$ deviating from (its normal
behaviour of) increasing linearly in time $t$; and commonly refers to
the power-law behaviour $\langle\Delta r^2(t)\rangle\sim t^c$ for
some $c\neq1$. Although the term ``anomalous'' diffusion was
originally coined to denote an anomaly --- in this case, a deviation
from normal diffusion --- anomalous diffusion has increasingly become
the norm \cite{rk}. Observed in many materials and systems such as in
fractal systems and disordered media \cite{bouchaud,havlin}, financial
markets \cite{stanley1}, transport in (crowded) cellular interiors
\cite{hofling1}, and migration of cells \cite{dietrich}, bacteria
\cite{ariel}, and animal foraging \cite{stanley2}, anomalous diffusion
has naturally received intense attention in the last decade. Interest
in the topic revolves largely around the following questions.  What
causes the exponent to differ from unity? Can one predict the exponent
from the underlying dynamics of the system? Are there universality
classes for systems exhibiting anomalous diffusion?

A number of distinct classes of stochastic processes have been
developed/identified for anomalous diffusion in the recent
years. Three most prominent theoretical (stochastic) models of
anomalous diffusion are:
\begin{itemize}
\item Transport on fractals: a popular model used for percolating and
  disordered materials \cite{deGennes,gefen,hofling2,havlin}, wherein
  the moving particle encounters obstacles on its path,
\item Continuous-time random walk (CTRW): a model where particles
  move from trap to trap \cite{scher,klafter1,klafter2,metz}, where 
  times of waiting at the traps as well as the trap-to-trap distance is
  power-law distributed, and
\item Gaussian models like fractional Brownian motion (fBm)
  which describes a Gaussian process with power-law memory
  \cite{kolm,mandelb}, attributed to the ``material medium'' that
  surrounds the particle that undergoes anomalous diffusion.
\end{itemize}
An overview of the available theoretical models, including a summary
of their distinctive features and stochastic properties can be found
in a recent perspective article \cite{metz_rev}.

Despite the above progress achieved, which model describes an instance
of (experimentally) observed anomalous diffusion is often the subject
of fierce debate, as evidenced by the recent case of anomalous
diffusion observed for tracer particles in cell cytoplasms
\cite{torre,selhuber,tejedor,jeon,golding,weber1}, where all three of
the above stochastic models have been fitted to the experimental data
\cite{mcguffee,saxton,selhuber,tejedor,jeon,golding,weber1,weber2}.
For physical systems where the dynamical rules for particles movement
are known (in contrast to a complicated medium like a cell cytoplasm),
one would expect to have a much easier task to model anomalous
diffusion, yet it can still remain quite a challenge. For polymeric
systems, where anomalous diffusion is commonplace, it is only recently
that one of us has established that the anomalous diffusion for tagged
monomers are explained by ``restoring forces'' that decay as a
power-law in time with the anomalous exponent of diffusion
\cite{panja1,panja1a}. From these characteristics it has been shown
that anomalous diffusion in polymeric systems can be modelled by a
Generalised Langevin Equation (GLE) with a memory kernel, and it
belongs to the class of fBm \cite{panja2}. The fBm characteristics of
anomalous diffusion have been verified for flexible
\cite{panja2a,maes} and semiflexible polymers \cite{kroy}, and polymer
membranes \cite{mem,popova,mizuochi}. Importantly, they have been used
to successfully explain the dynamics of translocation of polymers
across membranes \cite{panja3a,panja3b,panja3c,dubbel}. The fBm model
framework has been generalised/extended to the linear transport regime
for flexible polymers \cite{panja1}, and has similarly been used to
explain field-driven polymer translocation \cite{panja4} and polymer
adsorption \cite{panja5} for weak fields and adsorption energies. It
has also found applications in strong nonlinear regimes for flexible
polymers \cite{sakaue}.

In this paper, we take on characterising anomalous diffusion in
magnetisation space for the Ising model on a square lattice at the
critical temperature, undergoing Monte-Carlo spin-flip dynamics. That
the total magnetisation for this model exhibits anomalous diffusion
has been reported by one of us in Ref. \cite{walter}. Additionally, we
report that the magnetisation of a tagged line in the 2D Ising model,
and the magnetisation of a tagged plane in the 3D Ising model, also
exhibit anomalous diffusion. We argue that the anomalous diffusion for
all these quantities for the Ising model stems from time-dependent
restoring forces, decaying as power-laws in time --- with the
anomalous exponent of diffusion --- in striking similarity to
polymeric systems, and show that a closely analogous GLE formulation
holds for the Ising model as well. We obtain the memory kernel from
spin-spin correlation functions, and the formulation allows us to
consistently explain anomalous diffusion as well as anomalous response
of the Ising model to an externally applied magnetic field in a
consistent manner.

The organisation of this paper is as follows. In Sec. \ref{sec2} we
introduced the Ising model and report the anomalous exponents of
magnetisation. In Sec. \ref{sec3} we explain how restoring forces ---
that hold the key to anomalous diffusion --- develop and work. In
Sec. \ref{sec4} we develop the GLE formulation for anomalous diffusion
in the Ising model. The paper is concluded with a discussion
in Sec. \ref{sec5}.

\section{The anomalous diffusion in the Ising model at the critical temperature \label{sec2}}

\subsection{The model and dynamics \label{sec2a}}

We consider the Ising model on a square lattice. The Hamiltonian,
at zero external magnetic field, is given by 
\begin{equation}
  {\cal H}_0=-J\sum_{\langle ij\rangle}s_is_j,
  \label{a1} 
\end{equation} 
where $s_i = \pm 1$ is the spin at site $i$, and $J$ is the coupling
constant of interaction among the spins. The summation runs over all
the nearest-neighbour spins. The linear size of the system is $L$;
i.e., $0\le(i,j)<L$. Our samples satisfy periodic boundary conditions
at all times, and all properties we report here are studied (or
measured) at the critical temperature $T_c$.

The key quantity of focus in this paper is the mean-square
displacement (MSD) for magnetisation $M(t)$ at time $t$ as
\begin{equation} 
  \langle\Delta
  M^2(t)\rangle=\langle[M(t)-M(0)]^2\rangle,
  \label{a2} 
\end{equation}
where $M(t)$ can take several forms. All angular brackets in this
paper, including those in Eq. (\ref{a2}), denote ensemble average. In
the two-dimensional (2D) Ising model, we consider the respective cases
where it is the bulk magnetisation $M_{\text{2D},b}$, or the ``line
magnetisation'' $M_{\text{2D},l}$, the magnetisation of a tagged line
of spins in the $y$-direction. Similarly, in the three-dimensions, we
consider the bulk magnetisation $M_{\text{3D},b}$ and the
magnetisation $M_{\text{3D},p}$ of a tagged $xz$-plane.

We simulate the dynamics of the system using Monte Carlo moves,
following the Metropolis algorithm. At any time-step a spin is
randomly selected to flip, and the resulting change $\Delta E$, where
$E$ is the energy of the system, is measured. The move is accepted
with unit probability if $\Delta E\le0$; if not, then the move is
accepted with the usual Metropolis probability $e^{-\Delta E/(k_{\text
    B}T_c)}$, where $k_{\text B}$ is the Boltzmann constant.

All simulation results reported here use $k_B=J=1$.

\subsection{Anomalous diffusion in the Ising model \label{sec2b}}

Let us denote by $D$ the spatial dimension {\it of the support\/} of
the tagged magnetisation given by $M$, meaning $D=1$ for a tagged line
$D=2$ for bulk in the 2D Ising model, while for 3D Ising model $D=2$
for a tagged plane and $D=3$ for bulk. At short times $t\lesssim1$,
the individual spin flips in the model are uncorrelated, and since
there are $L^D$ spins all together in these entities spatial
dimensions, 
\begin{eqnarray}\langle\Delta M^2(t)\rangle\simeq
  L^Dt
  \label{shorttimes}.
\end{eqnarray}

At long times, $t\gg L^{z_c}$, where the $z_c$ is the dynamic
exponent for the Ising model at $T_c$, we expect $\langle
M(t)M(0)\rangle=0$. This means that
\begin{equation}
\begin{aligned}
\langle\Delta
M^2(t)\rangle\equiv\langle[M(t)-M(0)]^2\rangle\underrel{t\gg
  L^{z_c}}\quad=\,2\langle M^2\rangle,
\end{aligned}
\label{e2}
\end{equation}
which is a purely equilibrium quantity which we can calculate from the
equilibrium spin-spin correlations. We then have
\begin{equation}
  \langle M^2\rangle=\sum_{i\in L^D}\sum_{j\epsilon L^D}\langle
  s_i s_j\rangle=\sum_{i\in L^D}\sum_{j\epsilon
    L^D}r_{ij}^{2-d-\eta}\approx\int_1^L\frac{d^Dr}{r^{d-2+\eta}} \sim L^{2D-d+2-\eta},
  \label{eqM2}
\end{equation}
where $r_{ij}$ is the Euclidean distance between the two spins $i$ and
$j$, $d$ is the spatial dimension of the model (i.e., $d=2$ and 3 for
two- and three-dimensional Ising models respectively), and the
critical exponent $\eta$ is related to $\gamma$ and $\nu$ via the
scaling relation $2-\eta=\gamma/\nu$. (Note this result requires an
integral $\displaystyle{\int_1^L \frac{d^Dr}{r^{d-2+\eta}}}$ to be
dominated by large $r$, which is why we have excluded line
magnetisation $D=1$ in three dimensions $d=3$ from our paper.)

We now make the scaling assumption of an intervening power law with
time 
\begin{equation}
  \langle\Delta M^2(t)\rangle \propto t^c 
  \label{e3}
\end{equation}
connecting across intermediate times from Eq. (\ref{shorttimes}) at
$t\simeq 1$ to Eqs. (\ref{e2}-\ref{eqM2}) at $t\simeq L^{z_c}$. The
match at  $t\simeq 1$ forces  $\langle\Delta M^2(t)\rangle \simeq
L^D t^c$, and the match at large time $L^{z_c}$ then requires
$L^{D+c z_c} \simeq L^{2D-d+\gamma/\nu}$, leading to 
\begin{equation}
c=\frac{D-d+\gamma/\nu}{z_c}
\label{cfla}.
\end{equation}
The full scaling prediction valid for all $t\gg 1$ is then 
\begin{equation}
  \langle\Delta M^2(t)\rangle/L^{2D-d+\gamma/\nu}=f\left(t/L^{z_c}
  \right) 
  \label{fullscaling}
\end{equation}
where $f(x)\simeq x^c$ for $x\ll 1$. Using the values of the critical 
exponents corresponding to $k_B=J=1$, as presented in Table
\ref{tab2}, the explicit power laws for $1\lesssim t\lesssim L^{z_c}$
become 
\begin{equation}
 \langle \Delta M_{\text{2D},l}^2(t) \rangle\sim Lt^{(\gamma/\nu-1)/z_c}\approx  Lt^{0.35}
\nonumber
\end{equation}
\begin{equation}
 \langle \Delta M_{\text{2D},b}^2(t)\rangle\sim L^{2}t^{\gamma/(\nu z_c)}\approx L^2t^{0.81}
\nonumber
\end{equation}
\begin{equation}
 \langle \Delta M_{\text{3D},p}^2(t)\rangle\sim L^2t^{(\gamma/\nu-1)/z_c}\approx  L^2t^{0.48}.
\nonumber
\end{equation}
\begin{equation}
 \langle \Delta M_{\text{3D},b}^2(t)\rangle\sim L^3t^{\gamma/(\nu z_c)}\approx L^3t^{0.97},
\label{e4}
\end{equation}
indicating that anomalous diffusion in the Ising model is ubiquitous
at the critical temperature. As pointed out earlier, the results of
the bulk magnetisations were first obtained by one of us in
Ref. \cite{walter}.
\begin{table}[h]
\begin{tabular}{K{4.5cm}|K{2.5cm}|K{2.5cm}|K{2.5cm}|K{2.5cm}} 
\hline
Ising model dimension $d$& $\gamma$ & $\nu$ & $z_{c}$ & $T_{c}$\\
\hline
2& 7/4 & 1& 2.1665(12) & $\displaystyle{\frac{2}{\ln(1+\sqrt{2})}}$ \\
\hline
3& 1.237075(10) & 0.629971(4) & 2.03(4) & 4.5116174(2) \\
\hline
\end{tabular}
  \caption{The relevant critical exponents and the critical temperature in
  the Ising model \cite{binney,blote,landau}, using $k_B=J=1$ for the
  critical temperature $T_c$.\label{tab2}}
\end{table} 
\begin{figure*}
  \includegraphics[width=0.42\linewidth]{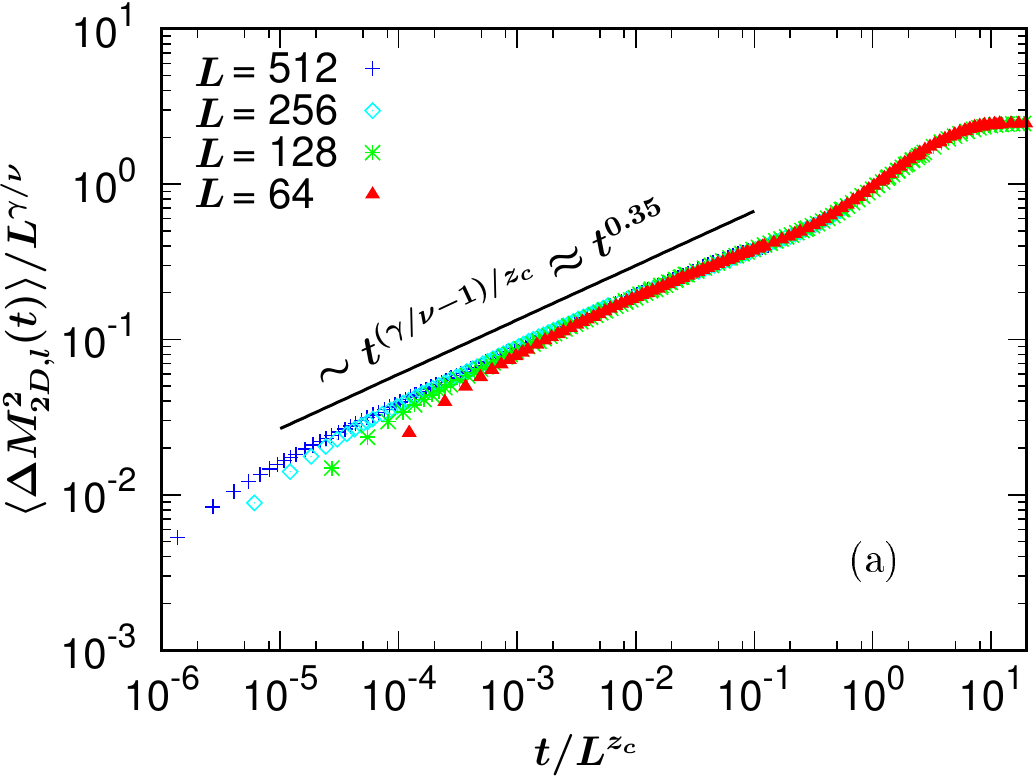}
  \hspace{5mm}
  \includegraphics[width=0.42\linewidth]{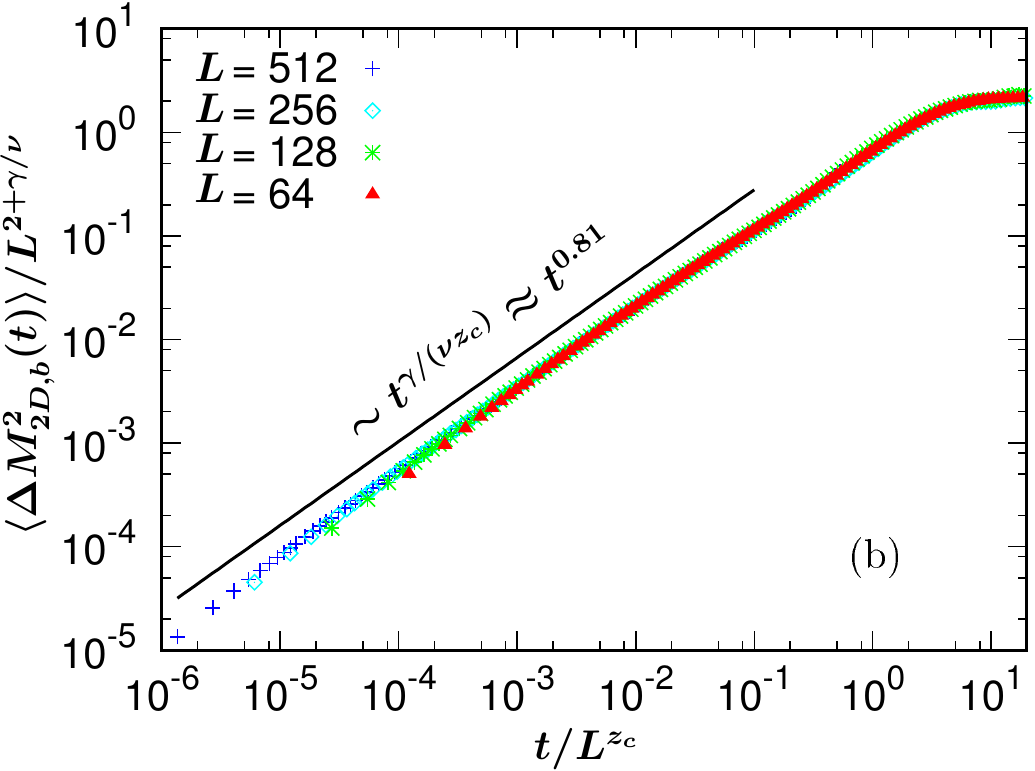}
  \includegraphics[width=0.42\linewidth]{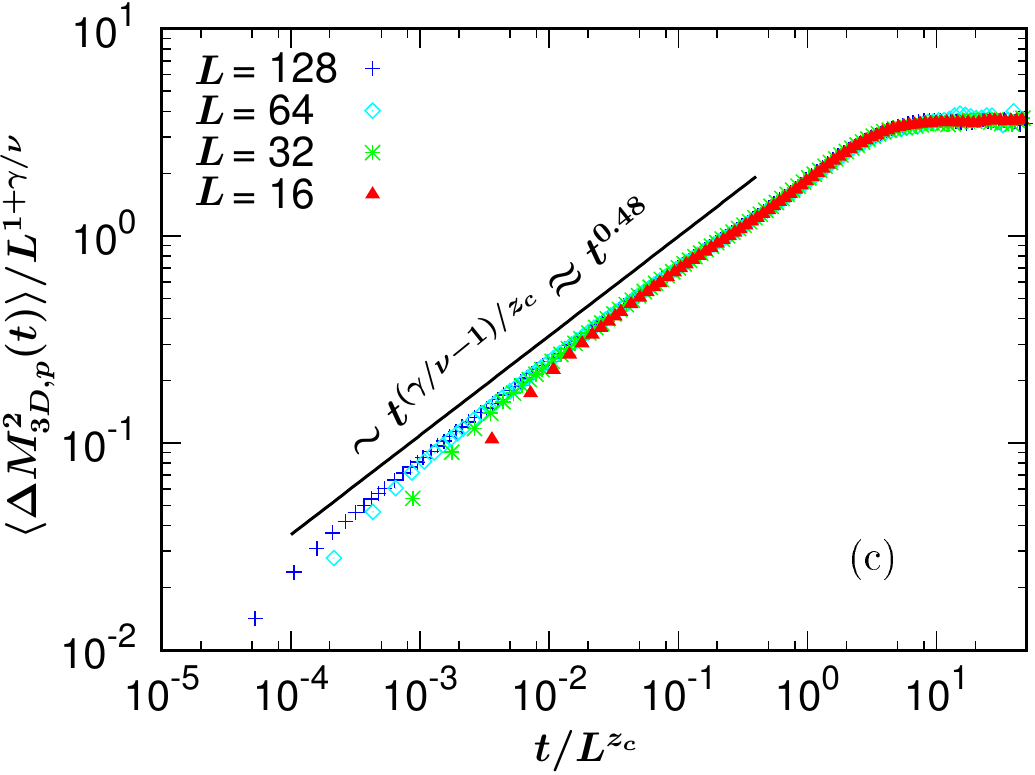}
  \hspace{5mm}
  \includegraphics[width=0.42\linewidth]{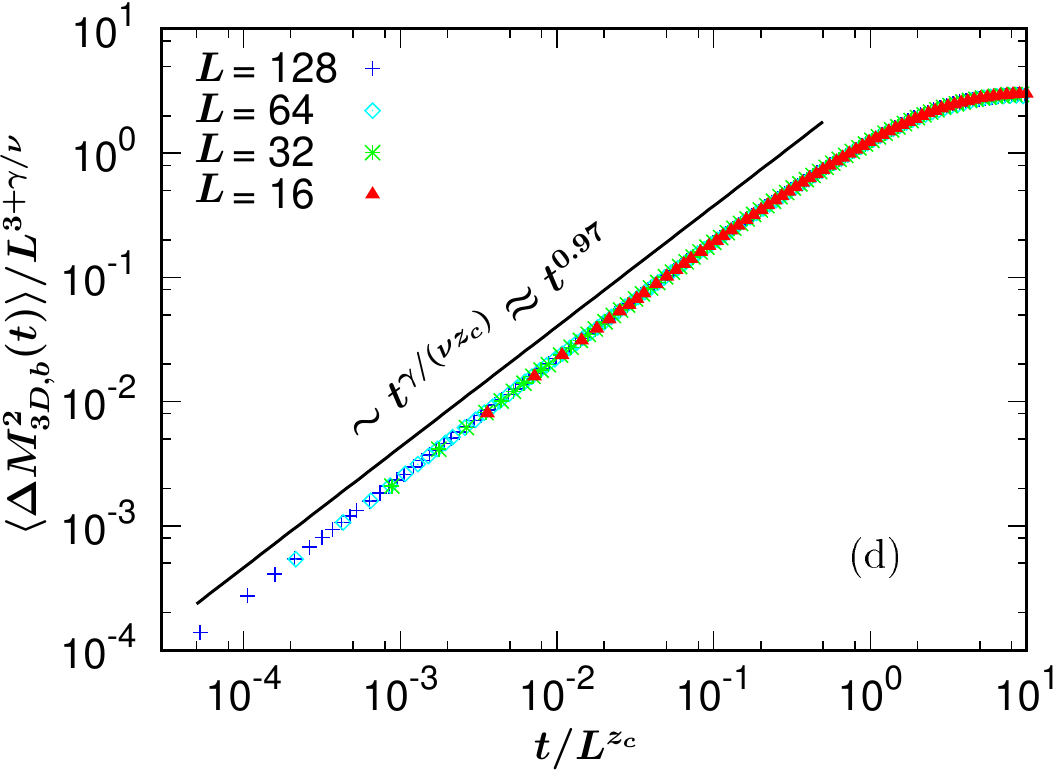}
  \caption{(colour online) The mean-square displacement (MSD) of the
    magnetisations $\langle \Delta M^2(t)\rangle$ in the Ising model
    at $T_c$: (a) tagged line magnetisation for 2D Ising model, (b)
    bulk magnetisation for 2D Ising model, (c) tagged plane
    magnetisation for 3D Ising model, and (d) bulk magnetisation for
    3D Ising model. The $x$- and $y$-axes are scaled according to 
    Eq. (\ref{fullscaling}) leading to excellent data collapse over
    different $L$. The black solid lines denote the power-laws shown
    in Eq. (\ref{e3}). \label{msd}}
\end{figure*}

The power-laws in Eq. (\ref{e4}) are verified in Fig. \ref{msd}. To
obtain these data, we first thermalised the system. We then produced a
number of independent time-series of $M(t)$, from which we measured
$\langle\Delta M^2(t)\rangle$. In some of the plots in Fig. \ref{msd}
we notice a small deviation from the power-laws at late times: we have
verified that this is caused by periodic boundary conditions --- they
are different when free boundary conditions are employed. Two examples
of this can be found in Appendix A.

\section{Restoring forces: the physics of anomalous diffusion in the
  Ising model \label{sec3}}
 
In this section we focus on the physics of anomalous diffusion. We
argue that anomalous diffusion in the Ising model stems from restoring
forces, in close parallel to polymeric systems.

\subsection{Restoring forces \label{sec3a}}

Imagine that the value of the tagged magnetisation $M$ changes by an
amount $\delta\!M$ due to thermal spin flips on the tagged line at
$t=0$. Due to the interactions dictated by the Hamiltonian, the spins
within and surrounding the tagged region, in the ensuing times, will
react to this change. This reaction will be manifest in the two
following ways: (a) the surrounding spins will to some extent adjust
to the change over time, and (b) during this time the value of $M$
will also readjust to the persisting values of the surrounding spins,
undoing at least a part of $\delta\!M$. It is the latter that we
interpret as the result of ``inertia'' of the surrounding spins that
resists changes in $M$, and the resistance itself acts as the
restoring force to the changes in the tagged magnetisation.
\begin{figure*}
  \begin{minipage}{0.47\linewidth}
    \includegraphics[width=\linewidth]{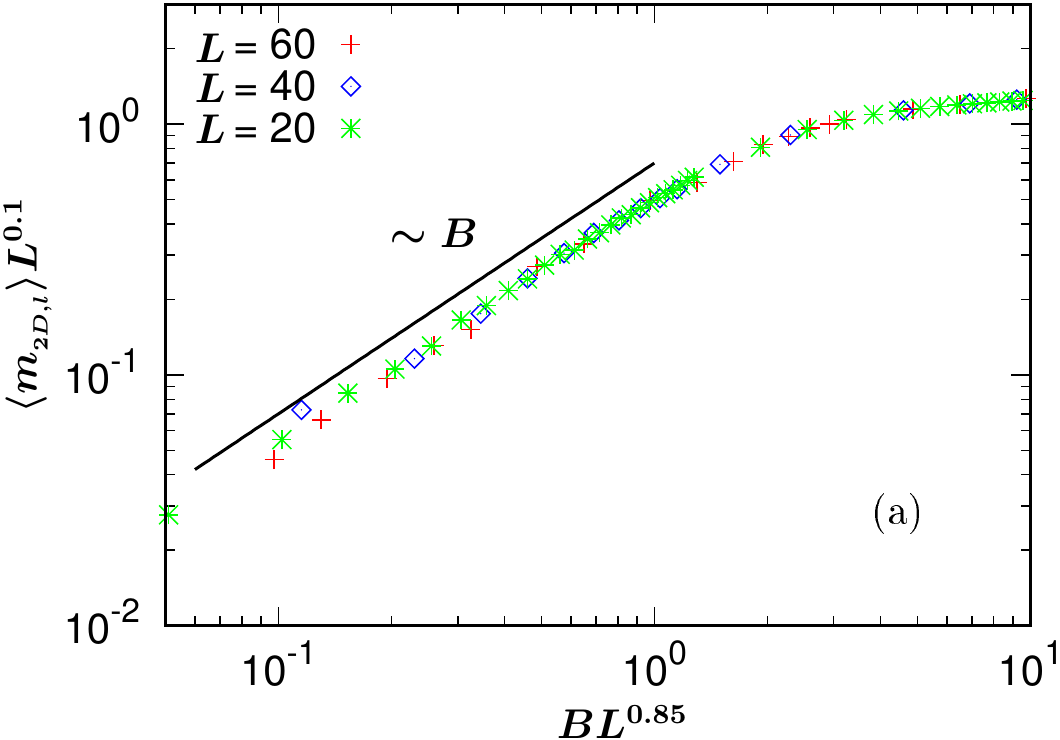}
  \end{minipage}
  \hspace{5mm}
  \begin{minipage}{0.47\linewidth}
    \includegraphics[width=\linewidth]{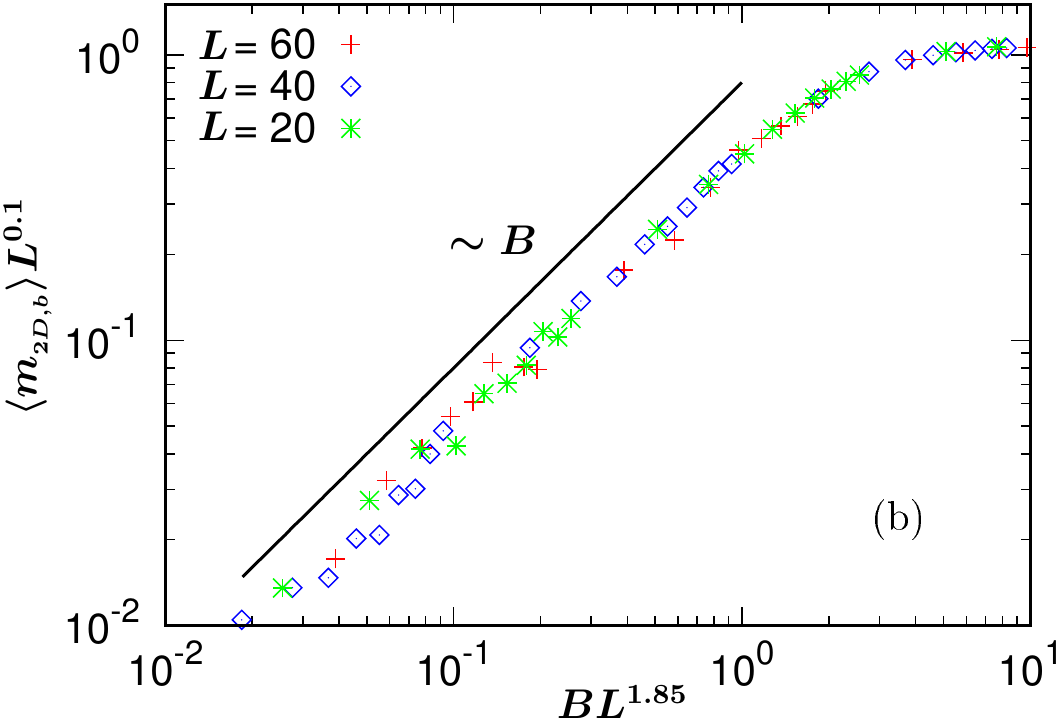}
  \end{minipage}
  \begin{minipage}{0.47\linewidth}
    \includegraphics[width=\linewidth]{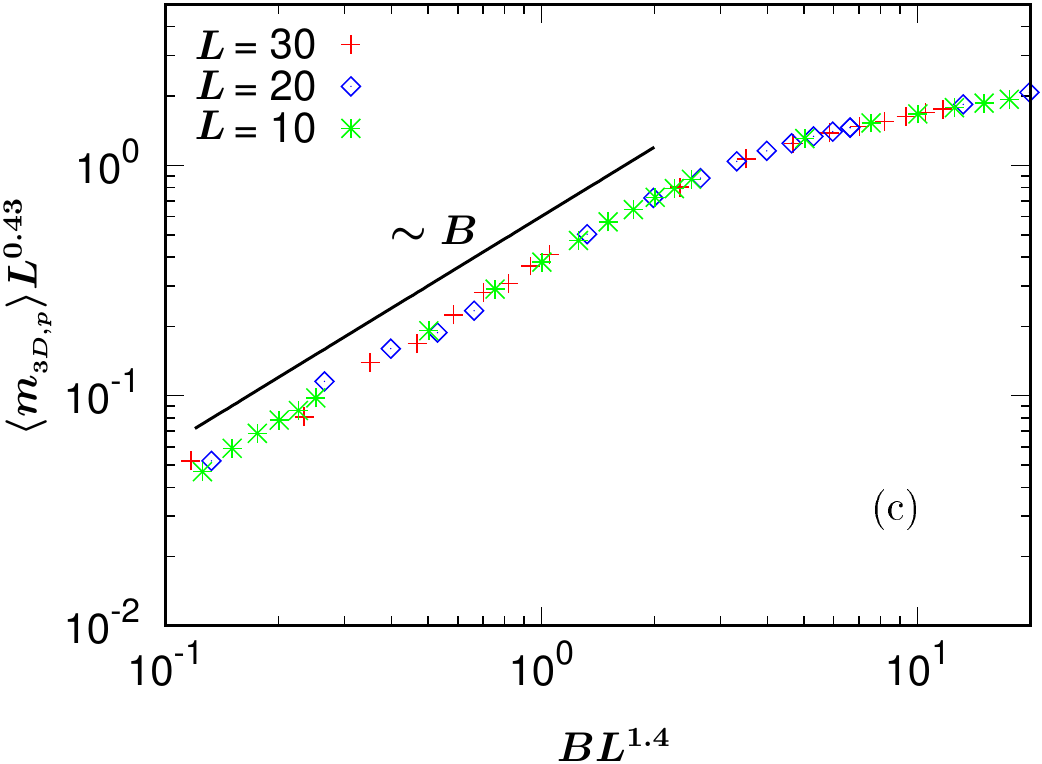}
  \end{minipage}
  \hspace{5mm}
  \begin{minipage}{0.47\linewidth}
    \includegraphics[width=\linewidth]{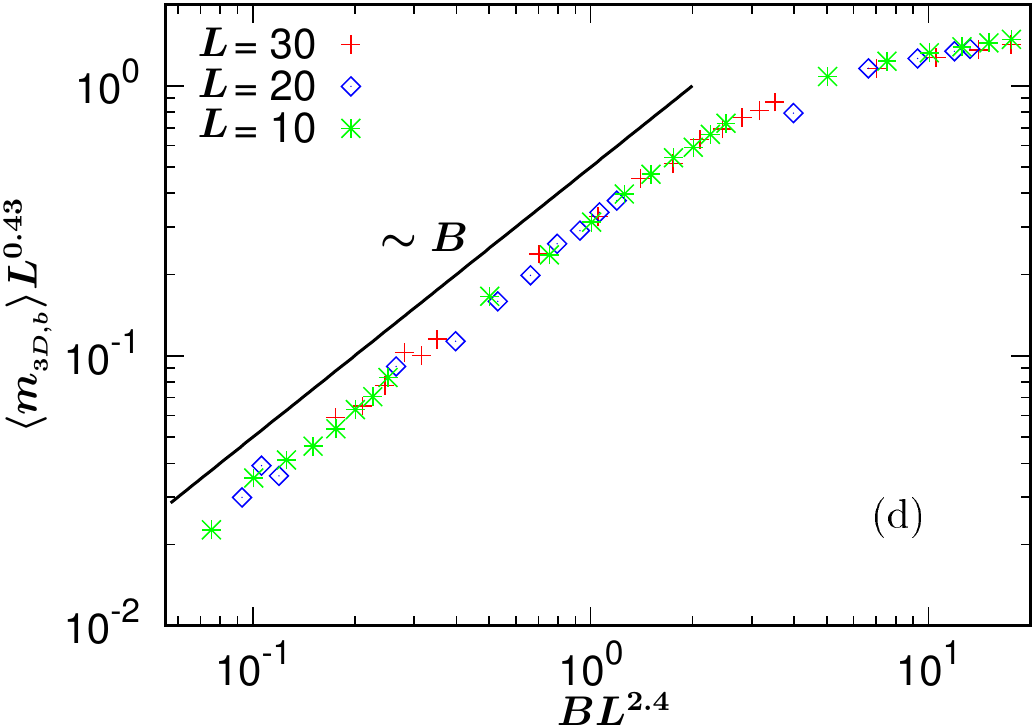}
  \end{minipage}
  \caption{(colour online) Plots showing the scaling form $\langle m
    \rangle L^\kappa \sim f(B L^\lambda)$ with
    $\kappa-\lambda=D-d+\gamma/\nu$, confirming Eq. (\ref{firstk2}).
    The (numerically found) values of $\lambda$ is $0.1$ in 2D and
    $0.43$ in 3D: (a) $\langle m_{\text{2D},l}\rangle$ (b) $\langle
    m_{\text{2D},b}\rangle$, (c) $\langle m_{\text{3D},p}\rangle$ and
    (d) $\langle m_{\text{3D},b}\rangle$ (note: $\gamma/\nu \approx
    1.75$ in 2D and $\approx 1.97$ in 3D).}
 \label{figappenda}
\end{figure*}

Since the part of the imposed change $\delta\!M$ will be partially
undone for $t>0$, we can expect the ``velocity'' autocorrelation
function $\langle \dot M(0) \dot M(t)\rangle$ to be negative, an
ingredient that we will use to establish the connection between the
restoring forces and anomalous diffusion in Sec. \ref{sec3c}.

\subsection{The time-decay behaviour of restoring forces
  \label{sec3b}}

The main ingredient to connect the restoring forces and anomalous
diffusion lies in how the former decays in time. To this end, we first
consider the following thought-experiment, along the line described
above in Sec. \ref{sec3a}. On an equilibrated set of samples of the
two-dimensional Ising model we create a small excess tagged
magnetisation $\delta\!M$ at $t=0$ with the constraint that we do not
allow this excess to be subsequently undone; this corresponds to an
imposed evolution of the tagged magnetisation
$d{M}(t)/dt=(\delta\!M)\,\delta(t)$, where $\delta(t)$ is the
Kronecker delta function. The resulting restoring force at later time
$t$ we will then write as
\begin{equation}
f(t)=-k(t)\,\delta\!M
\label{firstk1}
\end{equation}
where we interpret $k(t)$ as the magnetic analogue of a spring
constant: in conventional magnetic language this is related to the
susceptibility of the tagged magnetisation through $k^{-1}=L^D \chi$.

For long times $t\gg L^{z_c}$ our spring constant will be the
equilibrium one which is given by the equilibrium Fluctuation Theorem
as
\begin{equation}
  k^{-1} =\beta \langle M^2\rangle \sim L^{2D-d+\gamma/\nu}.
  \label{firstk2}
\end{equation}
Equation (\ref{firstk2}) can be confirmed by equilibrating samples
under the magnetic analogue of an externally applied force, which is
an external field applied to the tagged magnetisation (i.e., the field
is applied on the domain of support of the magnetisation), such that
the Hamiltonian becomes ${\cal H}={\cal H}_0-MB$. We then expect a
mean tagged magnetisation density $\langle
m\rangle=ML^{-D}=k^{-1}BL^{-D}\simeq B L^{D-d+\gamma/\nu}$ at small
values of $B$, which is the manifestation of linear response of the
system under weak external forcing. More generally, we can expect a
full scaling form $\langle m \rangle L^\kappa \sim f(B L^\lambda)$ for
some $\kappa$ and $\lambda$, where the scaling function $f(x)$ has the
property that $f(x\rightarrow\infty)\rightarrow$ constant, and
$f(x\rightarrow0)\sim x$ due to the linear dependence of $\langle
m\rangle$ on $B$ as $B\rightarrow0$. The latter condition implies that
$\kappa-\lambda=D-d+\gamma/\nu$.

The scaling form $\langle m\rangle=ML^{-D}=k^{-1}BL^{-D}\simeq B
L^{D-d+\gamma/\nu}$ with $\kappa-\lambda=D-d+\gamma/\nu$ is confirmed
in Fig. \ref{figappenda}. The quantity $\lambda$ is numerically found
to be 0.1 and 0.43 for Ising models in two- and in three-dimensions
respectively.
\begin{figure*}
  \begin{center}
    \includegraphics[width=0.4\linewidth]{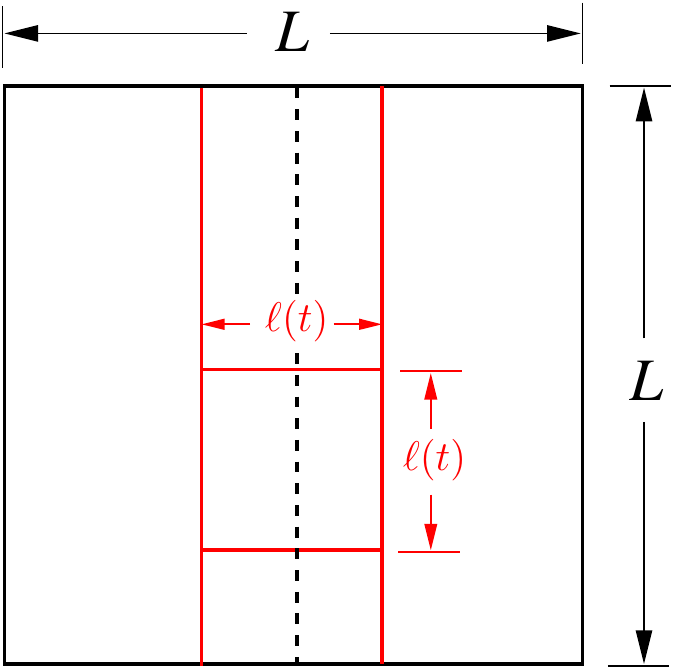}
    \caption{(colour online) The thought experiment performed on the
      tagged line magnetisation for the two-dimensional Ising model. A
      small excess (line) magnetisation $\delta\!M_{{\text{2D}},l}$ is
      created on the tagged line of spins, denoted by the dashed line,
      with the constraint that we do not allow the excess
      magnetisation to be undone. Up to time $t$, this action creates
      a rectangular zone of width $\ell (t)\sim t^{1/z_c}$ around the
      tagged line, shown by the red solid lines, which we can consider
      equilibrated to the new situation, in the following sense. If we
      consider the red square of size $\ell (t)\times\ell(t)$, then
      after time $t$ the spins therein will all have equilibrated to
      the segment of the tagged line within that square, and vice
      versa. \label{loft}}
  \end{center}
\end{figure*}

\begin{figure*}
  \includegraphics[width=.45\linewidth]{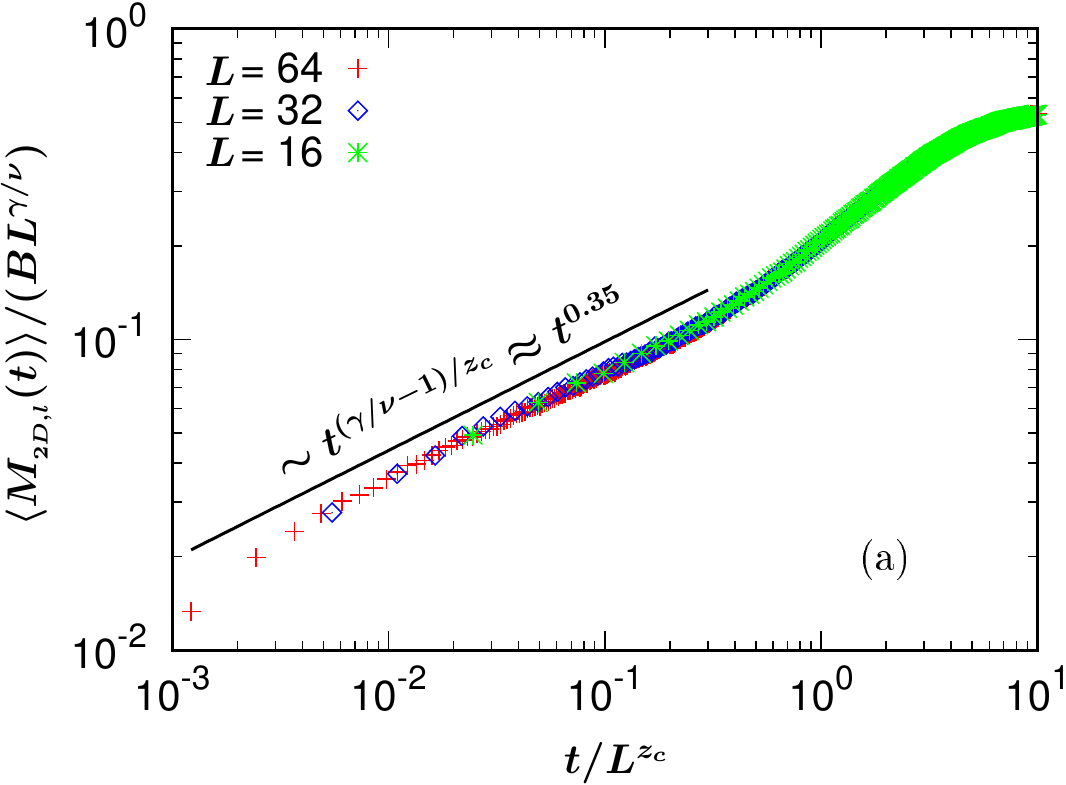}
  \includegraphics[width=.45\linewidth]{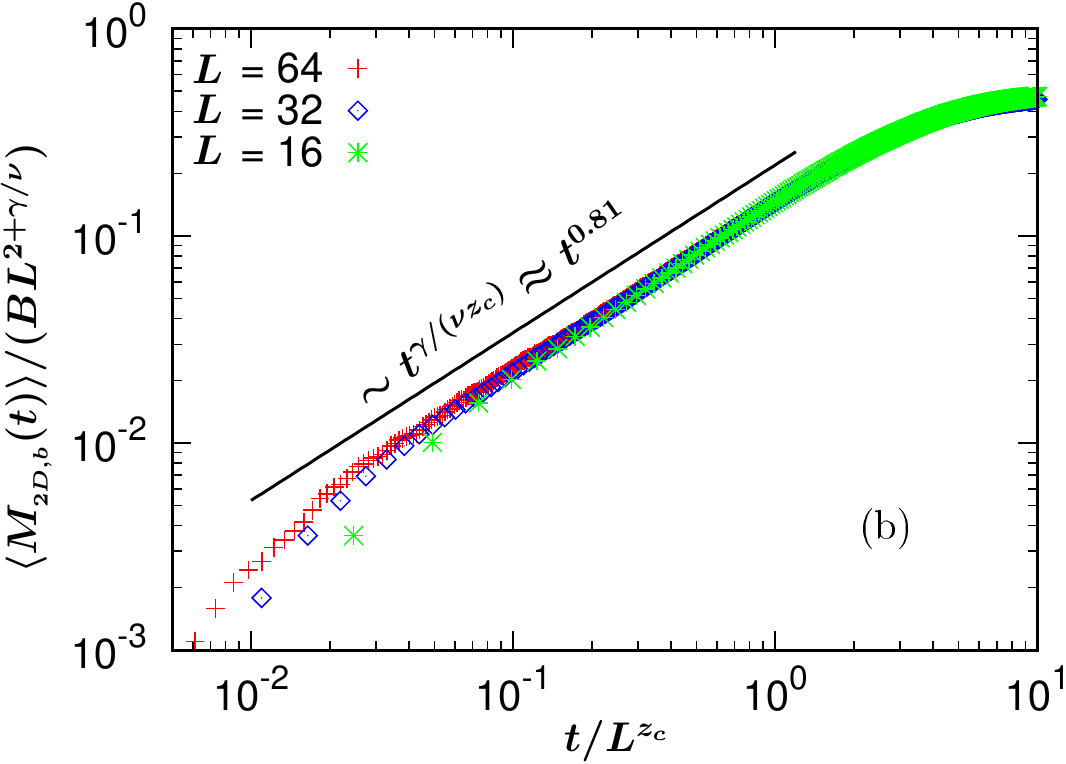}
  \includegraphics[width=.45\linewidth]{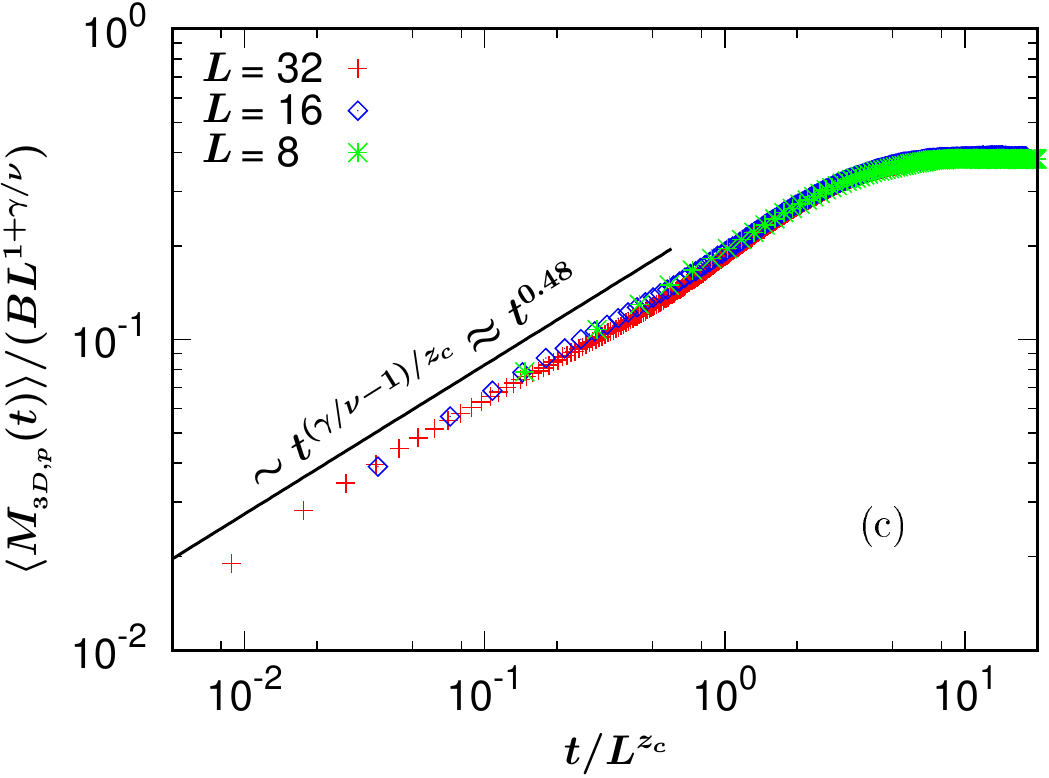}   
  \includegraphics[width=.45\linewidth]{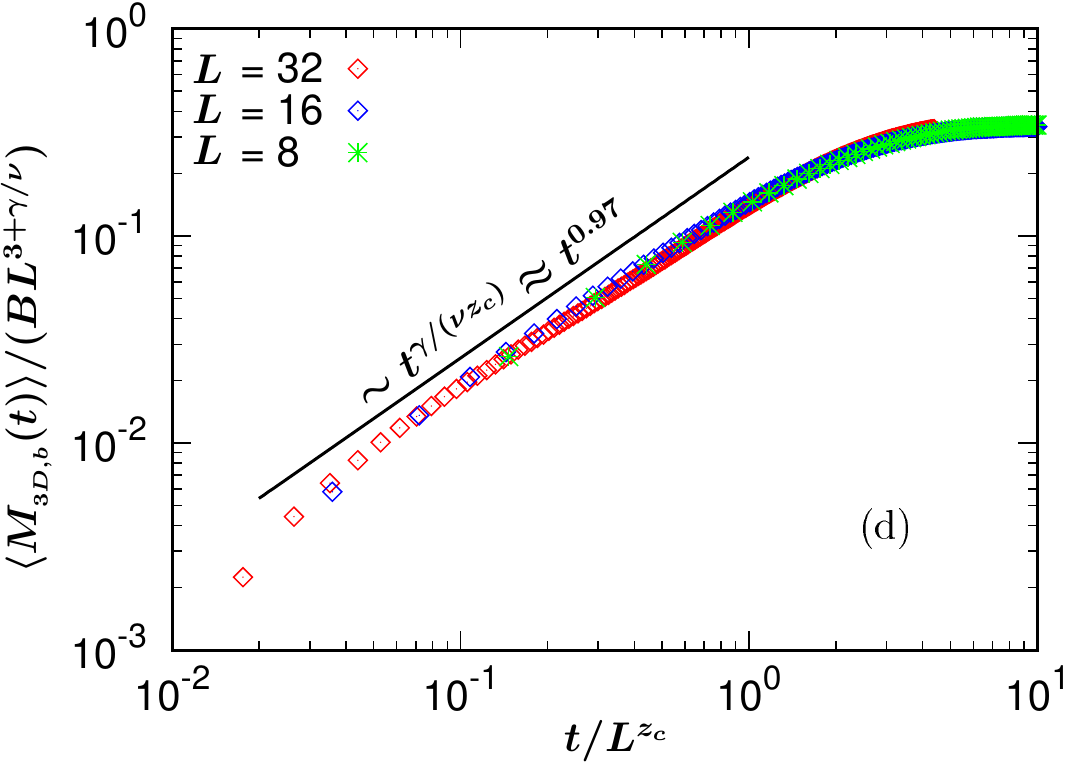}
  \caption{(colour online) Average magnetisations as a function of
    time when the magnetic field is switched on the equilibrated
    samples at $t=0$: (a) tagged line magnetisation $\langle
    M_{2D,l}(t)\rangle$ and (b) bulk magnetisation $\langle
    M_{2D,b}(t)\rangle$ for the 2D Ising model, and (c) the tagged
    plane magnetisation $\langle M_{3D,p}(t)\rangle$ and (d) bulk
    magnetisation $\langle M_{3D,b}(t)\rangle$ for the 3D Ising
    model. \label{Mt_vsB}}
\end{figure*}

For intermediate times we expect equilibrium response to be achieved
only locally across a length-scale $\ell (t)\sim t^{1/z_c}$ within and
around the tagged zone (see Fig. \ref{loft}). Within a region of the
tagged zone of side $\ell(t)$ we then expect a contribution of tagged
magnetisation
$\langle \Delta M\rangle_{\ell(t)}\sim B\ell(t)^{2D-d+\gamma/\nu}
$. Adding the response from $(L/\ell(t))^D$ such regions then leads to
\begin{equation}
\langle M(t)\rangle = k(t)^{-1} B \sim B L^D \ell(t)^{D-d+\gamma/\nu}
\sim B L^D t^c, 
\label{tiemedpendentk}
\end{equation}
where the exponent $c$ is as already given in Eq. (\ref{cfla}).  The
various cases of this result are verified in Fig. \ref{Mt_vsB}.

To summarise, the key result of this section is that if we create an
excess tagged magnetisation $\delta M$ at $t=0$ and do not allow it to
relax away, then a time-dependent restoring force $f(t)$ acts such as
would reverse it, where
\begin{eqnarray}
  \mbox{Tagged line magnetisation in two-dimensions:}\,\,f_{\text{rest}}(t)&=&-L^{-1}\,t^{-(\gamma/\nu-1)/z_c)}\,\delta\!M_{{\text{2D}},l},\nonumber\\
  \mbox{Bulk magnetisation in two-dimensions:}\,\,f_{\text{rest}}(t)&=&-L^{-2}\,t^{-\gamma/(\nu z_c)}\,\delta\!M_{{\text{2D}},b},\nonumber\\
  \mbox{Tagged plane magnetisation in three-dimensions:}\,\,f_{\text{rest}}(t)&=&-L^{-2}\,t^{-(\gamma/\nu-1)/z_c}\,\delta\!M_{{\text{3D}},p},\,\,\mbox{and}\nonumber\\
  \mbox{Bulk magnetisation in three-dimensions:}\,\,f_{\text{rest}}(t)&=&-L^{-3}\,t^{-\gamma/(\nu z_c)}\,\delta\!M_{{\text{3D}},b}.
\label{e3b7}
\end{eqnarray}

\subsection{Anomalous diffusion stems from these restoring forces \label{sec3c}} 
The main result of Sec. \ref{sec3b}, for which $\dot
M(t)\propto\delta(t)$, can be represented as the following formal
time-dependent ``impedance-admittance relation''
\cite{panja3a,panja3b,panja3c}
\begin{eqnarray}
  f_{\text{rest}}(t)=-\int_0^t dt'\,\mu(t-t') \dot M(t'),
  \label{e3c1}
\end{eqnarray}
with a causal memory function given by $\mu(t)=k(t) \sim L^{-D}t^{-c}$
as in Eq. (\ref{e3b7}) for $t>0$, and $\mu(t)=0$ for $t<0$. Equation
(\ref{e3c1}) is obtained from Eq. (\ref{e3b7}) using the superposition
principle: the total restoring force at time $t$ is a sum of all
preceding $\delta\!M$ values weighted by the (power-law) memory kernel
$\mu$. In this formulation, $\dot M(t)$ plays the role of current
through a circuit, with $f_{\text{rest}}(t)$ playing the role of the
voltage, and $\mu(t)$ is the time-dependent impedance. On the one
hand, this formulation means that $\langle f_{\text{rest}}(t)
f_{\text{rest}}(t')\rangle_{\dot M=0}=\mu(|t-t'|)$, while on the
other, we can invert Eq. (\ref{e3c1}) to express $\dot M(t)$ as a
function of $f_{\text{rest}}(t)$ involving the time-dependent
admittance $a(t)$ as
\begin{eqnarray}
  \dot M(t)=-\int_0^t dt'\,a(t-t') f_{\text{rest}}(t'),
  \label{e3c2}
\end{eqnarray}
and correspondingly $\langle \dot M(t)\dot
M(t')\rangle_{f_{\text{rest}=0}}=a(t-t')$, with the impedance and the
admittance following the relation $\tilde a(s)\tilde\mu(s)=1$ in the
Laplace space $s$. These imply that $a(t)=\langle \dot M(t)\dot
M(0)\rangle_{f_{\text{rest}=0}}\sim - L^D t^{c-2}$. Integrating this
quantity twice in time using the Green-Kubo relation we obtain
\begin{eqnarray}
  \langle\Delta M(t)\rangle^2\simeq L^D t^c 
  \label{e3c3}
\end{eqnarray}
 (we will return to this calculation more formally in
Sec. \ref{sec4}), leading us not only to the anomalous exponents of
Eq. (\ref{e3}), but also the correct $L$-dependent prefactors for the
data collapse in Fig. \ref{msd}. The results are summarised in Table
\ref{tab3}.
\begin{table*}
\begin{tabular}{K{3.5cm}|K{3.5cm}|K{3.5cm}} 
\hline\hline
Magnetisation of & $\mu(t)$ & $\langle\Delta M(t)\rangle^2$\\
\hline\hline
tagged line in 2D&$L^{-1}t^{-(\gamma/\nu-1)/z_c}$&$Lt^{(\gamma/\nu-1)/z_c}$\\
\hline
bulk in 2D&$L^{-2}t^{-\gamma/(\nu z_c)}$&$L^2t^{\gamma/(\nu z_c)}$\\
\hline
tagged plane in 3D&$L^{-2}t^{-(\gamma/\nu-1)/z_c}$&$L^2t^{(\gamma/\nu-1)/z_c}$\\
\hline
bulk in 3D&$L^{-3}t^{-\gamma/(\nu z_c)}$&$L^3t^{\gamma/(\nu z_c)}$\\
\hline 
\end{tabular} 
\caption{Memory functions and anomalous diffusion of magnetisation in
  the Ising model. The anomalous diffusion applies only until the
  terminal relaxation time scaling $\sim L^{z_c}$. \label{tab3}}
\end{table*}

\subsection{Restoring forces and anomalous diffusion: a similar story for
  polymer dynamics \label{sec3d}}

Although slightly off-topic, we now briefly point out that the
dynamics of the restoring forces and anomalous diffusion for
magnetisation in the Ising model is practically identical to those in
polymer dynamics
\cite{panja1a,panja1,panja2a,panja3a,panja3b,panja3c,panja4,panja5}. This
subsection forms the basis of Sec. \ref{sec4}, where we discuss the
Generalised Langevin Equation formulation of anomalous diffusion in
the Ising model.

Even though anomalous diffusion in polymeric systems is the norm rather
than an anomaly, we specifically pick the Rouse polymer to demonstrate
the similarity; for instance, the anomalous diffusion of a tagged
monomer in the Rouse model, which scales as $t^{2\nu/(1+2\nu)}$ until
the terminal Rouse time $\tau_R\sim N^{1+2\nu}$ (and diffusively
thereafter). Here, $\nu$ is the Flory exponent ($=3/4$ in two and
$\approx0.588$ in three dimensions), and $N$ is the polymer length.

Imagine that we move a tagged monomer by a small distance $\delta
\vec{r}$ at $t=0$ and hold it at its new position $\forall t>0$ (just
like in our thought experiment of Sec. \ref{sec3b}, where we created
an excess magnetisation $\delta M$ at $t=0$ and did not allow it to be
undone). For more details, we refer the reader to Ref. \cite{panja2a},
where we analysed this thought experiment. In the ensuing time, all
the monomers within a backbone distance $n_t\sim t^{1/(1+2\nu)}$,
counting away from the tagged monomer will equilibrate to the new
position of the tagged monomer. However, the end-to-end distance of
these equilibrated set of monomers is no longer their natural spatial
extent ($\sim n_t^\nu$), but is instead stretched by an amount
$\propto\delta\vec r$. With the (entropic) spring constant of these
$n_t$ equilibrated monomers scaling as $\sim n_t^{-2\nu}$, the mean
force the tagged monomer experiences at its new position is then given
by $\vec f_{\text{rest}}(t)\sim-n_t^{-2\nu}(\delta \vec{r})\sim
-t^{-2\nu/(1+2\nu)}(\delta \vec{r})$ [i.e., force=(spring
  constant)$\times$stretching distance]. This relation is identical in
formulation to Eqs. (\ref{e3b7}), and the rest of the
emulated analysis (\ref{e3c1}-\ref{e3c3}) leads one to the result that
the mean-square displacement of the tagged monomer increase as
$t^{2\nu/(1+2\nu)}$. Of course this result only holds till the
polymer's terminal Rouse time $\tau_R\sim N^{1+2\nu}$, just like the
anomalous diffusion in the Ising model survives until the terminal
relaxation time scaling $\sim L^{z_c}$.
 
The reader may find a comparison of Table I in Ref. \cite{panja1} and
Table \ref{tab3} of this paper interesting. Note that at the critical
temperature the system size $L$ corresponds to the polymer length $N$:
both systems reach criticality when these parameters reach infinity.

\section{Generalised Langevin Equation formulation for anomalous diffusion in the Ising model\label{sec4}} 

In the previous section we focused on the physics of the anomalous
diffusion in the Ising model. Using a thought experiment we argued
that the time-decay behaviour of the restoring forces is the key
ingredient to describe the relation between the restoring forces and
anomalous diffusion in terms of the memory function $\mu(t)$. Equation
(\ref{e3c1}) and its inverse formulation led us not only to the
anomalous exponents for the mean-square displacements, but also to the
correct $L$-dependent prefactors to obtain the data collapse in
Fig. \ref{msd}. These results pose now an interesting question: could
we formulate a stochastic differential equation for the anomalous
diffusion in the Ising model?

A comparison to the corresponding relations between the restoring
forces and anomalous diffusion for polymeric systems --- taken up in
the elaborate paper \cite{panja1} by one of us --- offers a clue to a
possible answer to the above question. Therein the (anomalous)
dynamics of a tagged monomer is shown to be described by the two
following stochastic differential equations involving the monomeric
velocity $v(t)$, the respective internal and external forces $f(t)$ and
$f_{\text{ext}}$ that it experiences, and the memory function
$\mu(t)$: \begin{eqnarray}
  \gamma v(t)&=&f(t)+q_1(t)\nonumber \\
  f(t)&=&-\int_0^tdt'\mu(t-t')\,v(t')+f_{\text{ext}}+q_2(t).
  \label{poleq}
\end{eqnarray}
Here $\gamma$ is the viscous drag on the monomer by the surrounding
(effective) medium, $q_1(t)$ and $q_2(t)$ are two noise terms
satisfying $\langle q_1(t)\rangle=\langle q_2(t)\rangle=0$, and the
fluctuation-dissipation theorems (FDTs) $\langle
q_1(t)\,q_1(t')\rangle\propto\gamma\delta(t-t')$ and $\langle
q_2(t)\,q_2(t')\rangle\propto\mu(t-t')$ respectively. (Note that factors of
$k_{\text B}T$ terms have been suppressed from these equations.) The
idea behind Eq. (\ref{poleq}) is that while the internal restoring
force builds on the history of the monomeric velocity, the latter
simply responds instantaneously to the force it experiences.

Similarity between the second one of Eq. (\ref{poleq}) and
Eq. (\ref{e3c1}) prompts us to propose the total force as
\begin{equation}
f(t)=-\int_0^t dt'\,\mu(t-t') \dot M(t')+f_{\text{ext}}+g(t),
\label{e4a1}
\end{equation}
for the Ising model, where $f_{\text{ext}}$ is simply the externally
applied force, such as a magnetic field. The noise term $g(t)$
satisfies the condition that $\langle g(t)\rangle=0$ and the
corresponding FDT $\langle g(t) g(t')\rangle=\mu(|t-t'|)$. As we have
done before, Eq. (\ref{e4a1}) can be inverted, in terms of the
admittance $a(t)$, to write
\begin{equation}
\dot M(t)=-\int_0^t dt'\,a(t-t') f(t')+\omega(t).
\label{e4a2}
\end{equation}
The noise term $\omega(t)$ similarly satisfies
$\langle\omega(t)\rangle=0$, and the FDT $\langle\omega(t)
\omega(t')\rangle=a(|t-t'|)$. The impedance and the admittance are
related to each other in the Laplace space as $\tilde
a(s)\tilde\mu(s)=1$. 

Additionally, we propose that in the Monte-Carlo dynamics, magnetisation
in the Ising model instantaneously responds to the internal force as
\begin{equation}
\zeta \dot{M}=f(t)+q(t),
\label{e4a3}
\end{equation}
with a damping coefficient $\zeta$ and a corresponding white noise term
$q(t)$. Thereafter, having combined Eqs. (\ref{e4a1}) and (\ref{e4a3})
we obtain
\begin{equation}
\zeta \dot{M}=-\int_0^t dt'\,\mu(t-t') \dot{M}(t')+f_{\text{ext}}+g(t)+q(t),
\label{e4a4}
\end{equation}
or
\begin{equation}
 \dot{M}=\int_0^t dt'\,\theta(t-t')[f_{\text{ext}}+g(t')+q(t')],
\label{e4a5}
\end{equation}
where in the Laplace space $\tilde
\theta(s)[\zeta+\tilde\mu(s)]=1$.Here, without the $\zeta$ term
$\theta(t)$ is identical to $a(t)$, introduced in Eq. (\ref{e3c2}).

At zero external magnetic field the dynamics of $M$ simplifies to
\begin{equation}
 \dot{M}=\int_0^t dt'\,\theta(t-t')[g(t')+q(t')],
\label{e4a5a}
\end{equation}
similar to Eq. (\ref{poleq}) for polymeric systems. Without further
ado, we then simply follow Ref. \cite{panja1} to conclude, with
$\mu(t)\sim L^{-D}t^{-c}$, that
\begin{equation}
\langle \dot M(t) \dot M(t') \rangle=-\theta(t-t')\sim -L^D (t-t')^{c-2}.
\label{e4a6}
\end{equation}
Note that in Eq. (\ref{e4a6}) we have ignored the $\zeta$ term, which
essentially means that we are ingoring the (uninteresting) time-scale
$\lesssim\zeta^{-1}$. Subsequently, by integrating the
Eq. (\ref{e4a6}) twice in time using the Green-Kubo relation, the MSD
of the magnetisation can be obtained as
\begin{equation}
\langle\Delta M^2(t)\rangle\sim L^D t^c,
\label{e4a7}
\end{equation}
which are the same results obtained in Eq. (\ref{e3c3}).  An example
verification for the velocity autocorrelation function (\ref{e4a6})
can be found in Appendix B.

This GLE formulation demonstrates that the anomalous diffusion in the
Ising model at the critical temperature is non-Markovian, with a
power-law memory function $\mu(t)$. Quite simply, if $\mu(t)\sim
t^{-c}$, then the anomalous diffusion exponent is $c$.

\subsection{Numerical confirmation of the GLE formulation (and determination of the damping coefficient $\zeta$)}

It is now imperative that we numerically test our proposed GLE
formulation for anomalous diffusion for the Ising model. Our key test
is to check the FDT
$\langle
f_{\text{rest}}(t)f_{\text{rest}}(t')\rangle_{\dot{M}=0}=\mu(t-t')$,
for which we describe our approach below, followed by presentation of
the numerical results.

Conceptually, the task is simple. At a fixed value of $M$, i.e.,
$\dot M=0$ at all times, we need to numerically measure
$\langle f_{\text{rest}}(t) f_{\text{rest}}(t')\rangle$. However, we
cannot measure forces in the Monte Carlo dynamics of the model since
by definition one does not have forces in discrete lattice models. In
order to circumvent this difficulty, we use Eq. (\ref{e4a3}) as a
proxy for $f(t)$ by choosing $\zeta=1$ and use the value
$\overline{\dot M}_{\text{free}}$ (see below), which would have
applied to the tagged magnetisation if the fixed $M$ constraint were
to be lifted at that time.

We start with a thermalised system at $t=0$. For $t>0$ we fix the
value of $M$ (this does not mean that all tagged spins are frozen),
which we achieve by performing non-local spin-exchange
moves. Specifically, for the magnetization of a tagged line in 2D and
tagged plane in 3D, we avoid extreme values of $M$ by choosing to fix
it in the interval $-0.2<m=ML^{-D}<0.2$ (note that in the scaling
limit all values of $m$ belong to this range). We then keep taking
snapshots of the system at regular intervals, and compute, at every
snapshot (denoted by $t$), the expectation value $\overline{\dot
  M}_{\text{free}}(t)$ conditional on the current configuration, which
for our Metropolis Monte-Carlo dynamics is given by,
\begin{eqnarray}
  \overline{\dot M}_\text{free}(t)=\sum_{i\,\in\,{\text{tagged}}}(-2s_i)\,\text{Min}\left(1,e^{-\Delta E_i/(k_{\text B}T_c)}\right)=f(t).
  \label{spinflip}
  \end {eqnarray}
This means that for every snapshot we take, we consider an attempt
to flip each spin in turn and find the expected change in $M$ which
would have occurred if this move had been implemented, totalled over
all the spins.

\begin{figure*}
  \begin{minipage}{0.47\linewidth}
    \includegraphics[width=\linewidth]{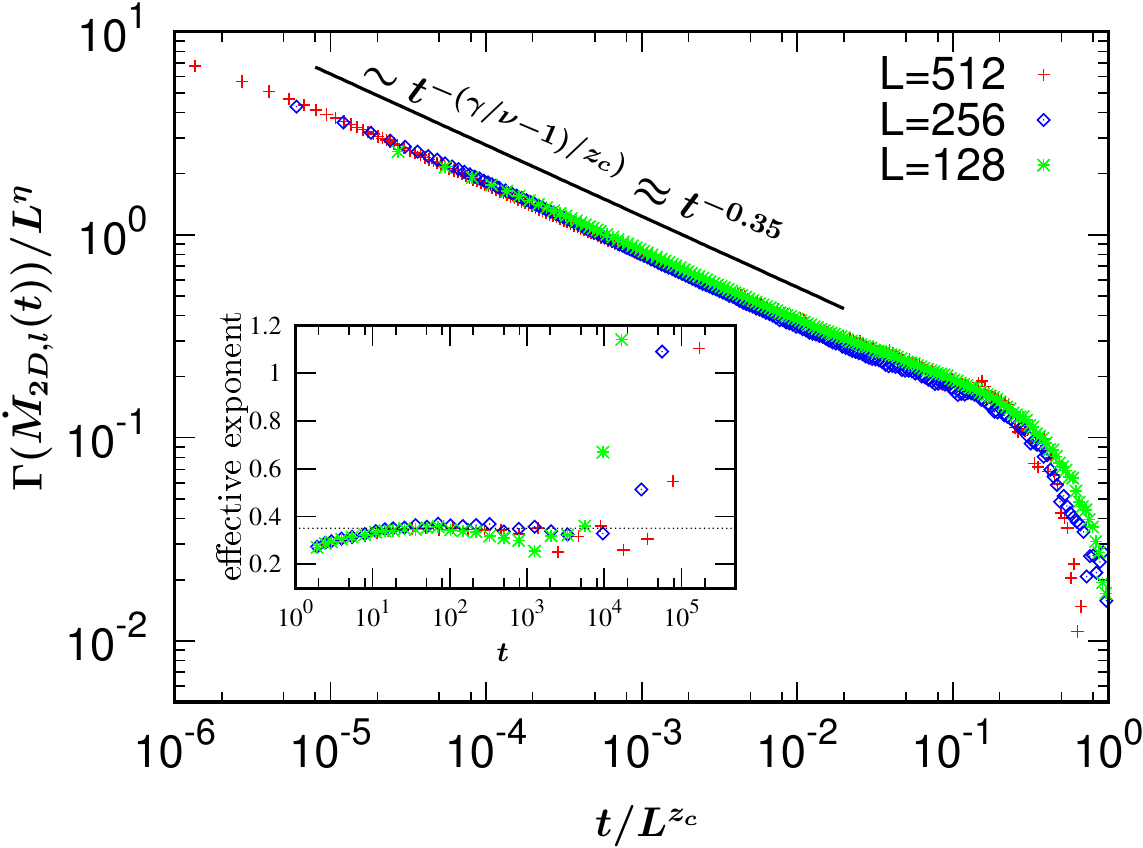}
  \end{minipage}
  \hspace{5mm}
  \begin{minipage}{0.47\linewidth}
    \includegraphics[width=\linewidth]{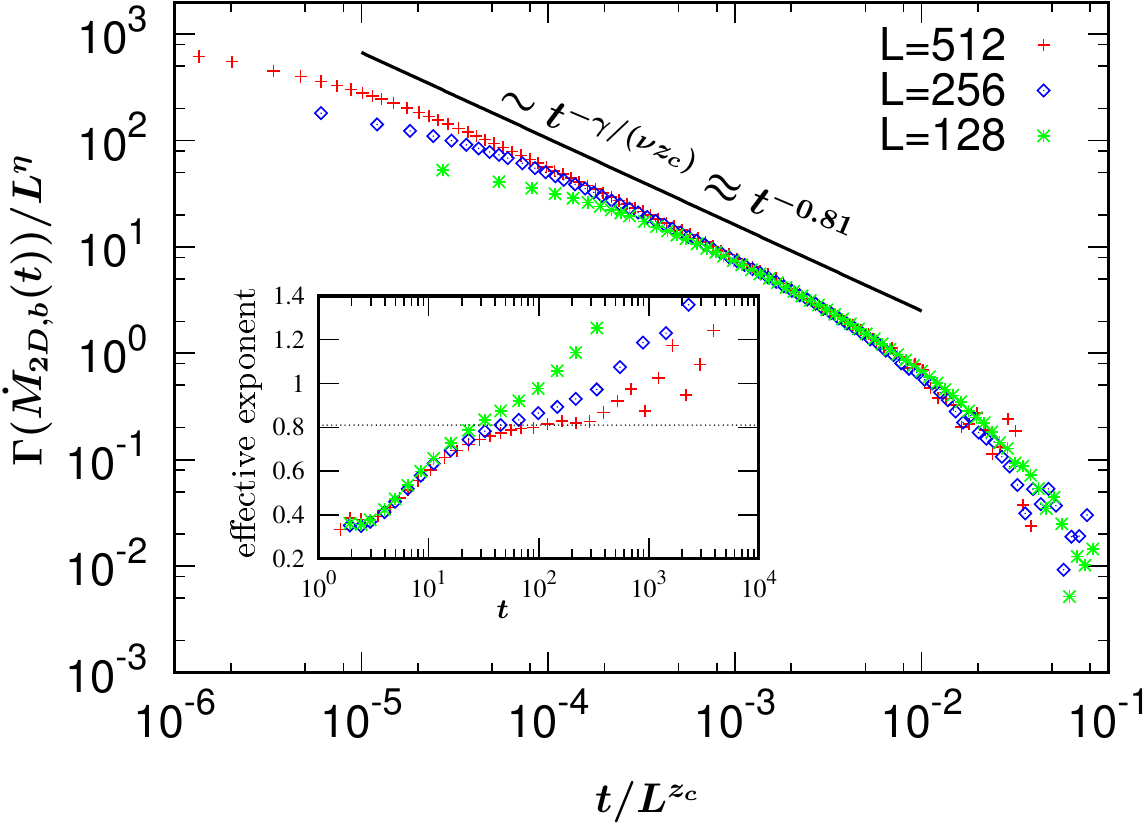}
  \end{minipage}
  \begin{minipage}{0.47\linewidth}
    \includegraphics[width=\linewidth]{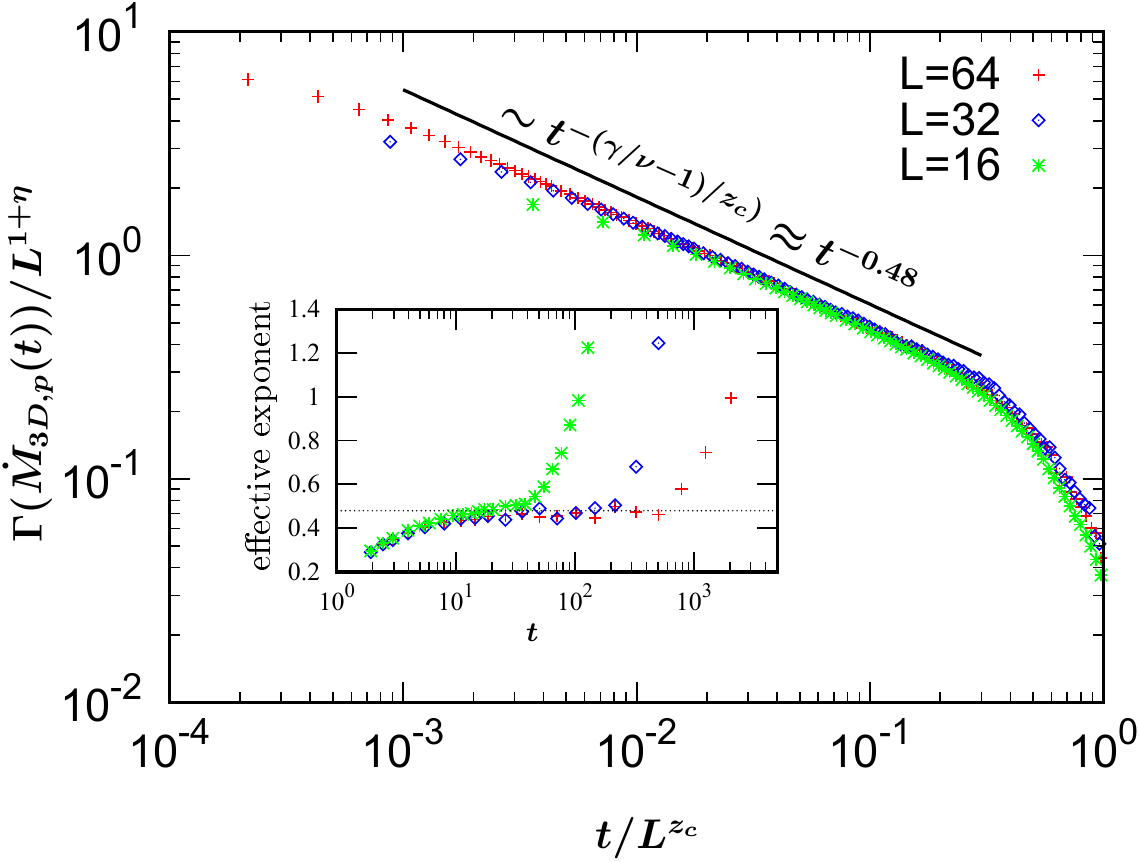}
  \end{minipage}
  \hspace{5mm}
  \begin{minipage}{0.47\linewidth}
    \includegraphics[width=\linewidth]{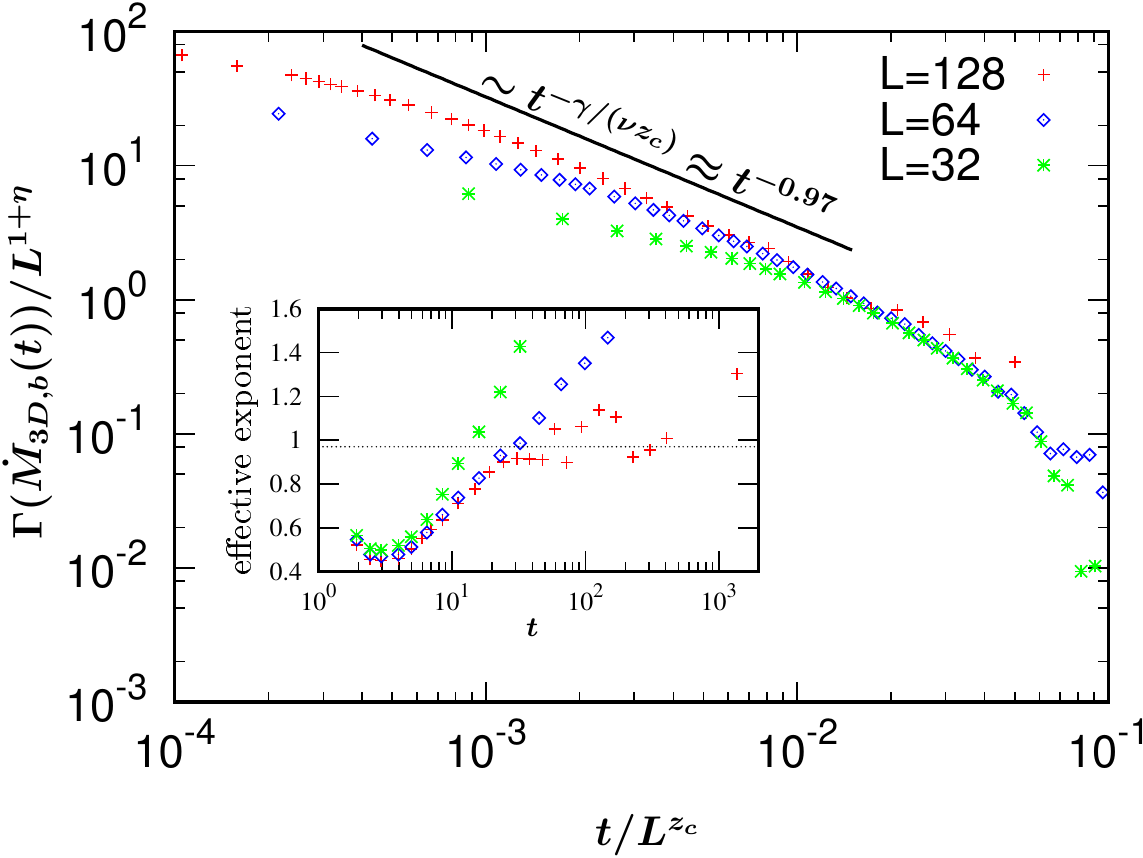}
  \end{minipage}
  \caption{(colour online) Plots showing the scaling of (a)
    $\Gamma(\dot{M}_{\text{2D},l}(t))$, (b)
    $\Gamma(\dot{M}_{2D,b}(t))$, (c) $\Gamma(\dot{M}_{3D,p}(t))$ and
    (d) $\Gamma(\dot{M}_{3D,b}(t))$ as a function of $t / L^{z_c}$.
    Insets show the effective exponent, numerically obtained
    derivative $-d(\ln\Gamma)/d(\ln t)$, with the dotted lines denoting
    the expected values of the slope. Note that the critical exponent
    $\eta$ is related to $\gamma$ and $\nu$ via the scaling relation
    $\eta=2-\gamma/\nu$.\label{e4f1}}
\end{figure*}

Finally, we note that since simulations are performed for finite
systems with $M$ fixed at its $t=0$ value, in any particular run we
need a non-zero value of $f_{\text{ext}}=-\langle f(t)\rangle$ acting
to sustain the initial value of $M$. Further, given that that in our
proxy measurement for $f(t)$ using Eq. (\ref{e4a3}) we can only access
$f_{\text{rest}}(t)+f_{\text{ext}}$, but not $f_{\text{rest}}(t)$
directly, it is the quantity 
\begin{equation}
\begin{aligned}
\Gamma(\dot{M}_{\text{2D},l}(t)) &= \langle \dot{M}(t)\dot{M}(t')\rangle-\langle
  \dot{M}(t)\rangle\langle \dot{M}(t')\rangle \\
  &=L^{2D}(\langle f(t)f(t')\rangle-\langle
  f(t)\rangle\langle f(t')\rangle)
  \label{ffcorr}
\end{aligned}
\end{equation}
that should correctly proxy
$\langle g(t) g(t') \rangle_{\dot{M}=0}=\mu(t-t')$, and we expect the
following results:
\begin{eqnarray}
  \Gamma(\dot{M}_{\text{2D},l}(t))\sim Lt^{-(\gamma/\nu-1)/z_c}\approx Lt^{-0.35},\nonumber\\
  \Gamma(\dot{M}_{\text{2D},b}(t))\sim  L^2t^{-\gamma/(\nu z_c)}\approx L^2t^{-0.81},\,\,\nonumber\\
  \Gamma(\dot{M}_{\text{3D},p}(t))\sim L^2t^{-(\gamma/\nu-1)/z_c}\approx L^2t^{-0.48},\,\,\nonumber\\
  \Gamma(\dot{M}_{\text{3D},b}(t))\sim L^3t^{-\gamma/(\nu z_c)}\approx L^3t^{-0.97}.
  \label{e4b2}
\end{eqnarray}
These results are verified in Fig. \ref{e4f1}, along with the
effective exponents as numerically obtained derivative
$d(\ln\Gamma)/d(\ln t)$ as insets.

In Fig. {\ref{e4f1}}, the data quality for 3D bulk at long times
suffers from the difficulty of collecting statistically independent
datasets at long times. There are also small deviations from the
power-laws at late times for line magnetisation in 2D and plane
magnetisation in 3D; we suspect that these relate to similar
deviations observed in Fig. \ref{msd}.
\begin{table*}
\begin{tabular}{K{6cm}|K{6.5cm}|K{3cm}} 
\hline\hline
System & Estimated exponent from Fig. \ref{e4f1}&Expected value\\
\hline\hline
tagged line in 2D, $L=512$&$-0.35\pm 0.02$&$-0.35$\\
\hline
bulk in 2D, $L=512$&$-0.81\pm 0.03$&$-0.81$\\
\hline
tagged plane in 3D, $L=64$&$-0.47\pm 0.02$&$-0.48$\\
\hline
bulk in 3D, $L=128$&$-0.97\pm 0.04$&$-0.97$\\
\hline 
\end{tabular} 
\caption{Power-law exponents from data in Fig. \ref{e4f1}, together
  with error bars, for the respective largest system sizes. Evidently,
  the data compare well with the expected exponents. \label{tab4}}
\end{table*}

Additionally, we have followed the procedure described in
Ref. \cite{goldstein} to obtain the power-law exponents from data in
Fig. \ref{e4f1}; these values, together with the error bars, for the
respective largest system sizes, can be found in Table \ref{tab4}. We
have chosen the largest system sizes for this purpose since they
contain the least amount of finite size effects.

\subsection{The GLE formulation for driven Ising systems}

The GLE formulation (\ref{e4a2}-\ref{e4a3}) also describes the
anomalous response of the model to external magnetic fields. Starting
from Eq. (\ref{e4a5}) and focusing on the response to an external
field $f_{\text{ext}}=B$ switched on at $t=0$, we readily obtain the
results of Eq. (\ref{tiemedpendentk}) for the tagged magnetisation
induced for times $1\lesssim t\lesssim L^{z_c}$ by taking an ensemble
average that reduces the noise terms $g(t)$ and $q(t)$ to zero;
specifically, 
\begin{eqnarray}
  \langle M_{\text{2D},l}(t)\rangle\sim BLt^{(\gamma/\nu-1)/z_c}\approx BLt^{0.35},\nonumber\\
  \langle M_{\text{2D},b}(t)\rangle\sim BL^2t^{\gamma/(\nu z_c)}\approx BL^2t^{0.81},\,\,\nonumber\\
 \langle M_{\text{3D},p}(t)\rangle\sim BL^2t^{(\gamma/\nu-1)/ z_c}\approx BL^2t^{0.48},\,\,\nonumber\\
\langle M_{\text{3D},b}(t)\rangle\sim BL^3t^{\gamma/(\nu z_c)}\approx BL^3t^{0.97},
 \label{e4c5}
\end{eqnarray}
which have been verified already in Fig. \ref{Mt_vsB}.

\section{Discussion\label{sec5}}

In summary, in this paper we report that the Ising model in two and
three dimensions exhibit ubiquitous anomalous diffusion behaviour at
the critical temperature. We have performed four case studies for
this: the bulk magnetisations, magnetisation of a tagged line in 2D
and that of a tagged plane in 3D. We have argued that the anomalous
diffusion stems from a time-dependent restoring force that involves a
power-law memory kernel. We have derived these power-laws as well as
the corresponding $L$-dependent prefactors.

Further, we have shown that the physics of anomalous diffusion in the
Ising model bears strong similarities to that in polymeric systems,
allowing us to propose a GLE description for anomalous diffusion in
the Ising model. We have also verified that the anomalous diffusion
for the tagged magnetisations in the Ising model belongs to the
fractional Brownian motion (fBm) class, although we do not explicitly
report it in this paper. We have numerically tested the specific
aspects of the GLE (such as the FDTs), and the GLE description is also
consistent with the observed anomalous response of magnetisations to
externally applied magnetic fields. In a future paper, work on which
is already in progress, we will expand the GLE formulation to the
Ising model around the critical temperature.

Having said the above, we have
not mathematically proved the GLE, neither the fBm, for the Ising
model. Some other kinds of models may also be consistent with the
anomalous diffusion behavior observed by us in this paper. They
should, however, feature restoring forces, transient response to an external magnetic field, and a negative velocity autocorrelation function
(observed in Fig. \ref{velauto}), in a consistent manner as presented here. In particular, we note that the Ising model we study here is
at equilibrium at $T_c$, and therefore time-reversible, so anomalous
diffusion models that are developed for time-irreversible aging-type
systems will not be applicable here.

Finally, we believe that the anomalous diffusion of the order
parameter at the critical temperature can be found in other Ising-like
systems, and if so, the GLE formulation introduced in this paper can
be employed to describe those anomalous behaviour as well. In
particular, if we know the critical temperature $T_c$, the critical
exponents $\gamma$ and $\nu$ for a specific Ising-like system, then
this method can be used to obtain the critical dynamical exponent
$z_c$ from the power-laws as well as the scaling of the terminal time
$\sim L^{z_c}$ (in other words, anomalous diffusion can be effectively
used to measure the critical dynamic exponent $z_c$). We will test
these ideas in our future work.

\section*{Acknowlegement \label{sec6}}
W. Z. acknowledges financial support from the CSC (Chinese Scholarship Council).

\section*{Appendix A \label{sec7a}}

\setcounter{figure}{0}
\renewcommand{\thefigure}{A\arabic{figure}}

\begin{figure*}
  \includegraphics[width=0.42\linewidth]{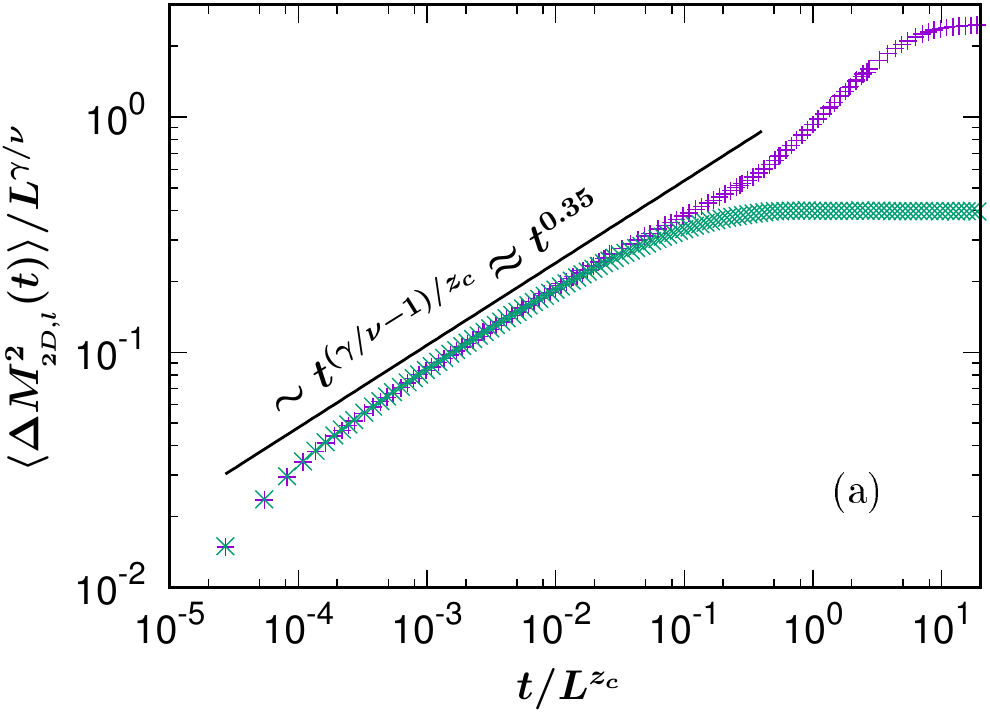}
  \hspace{5mm}
  \includegraphics[width=0.42\linewidth]{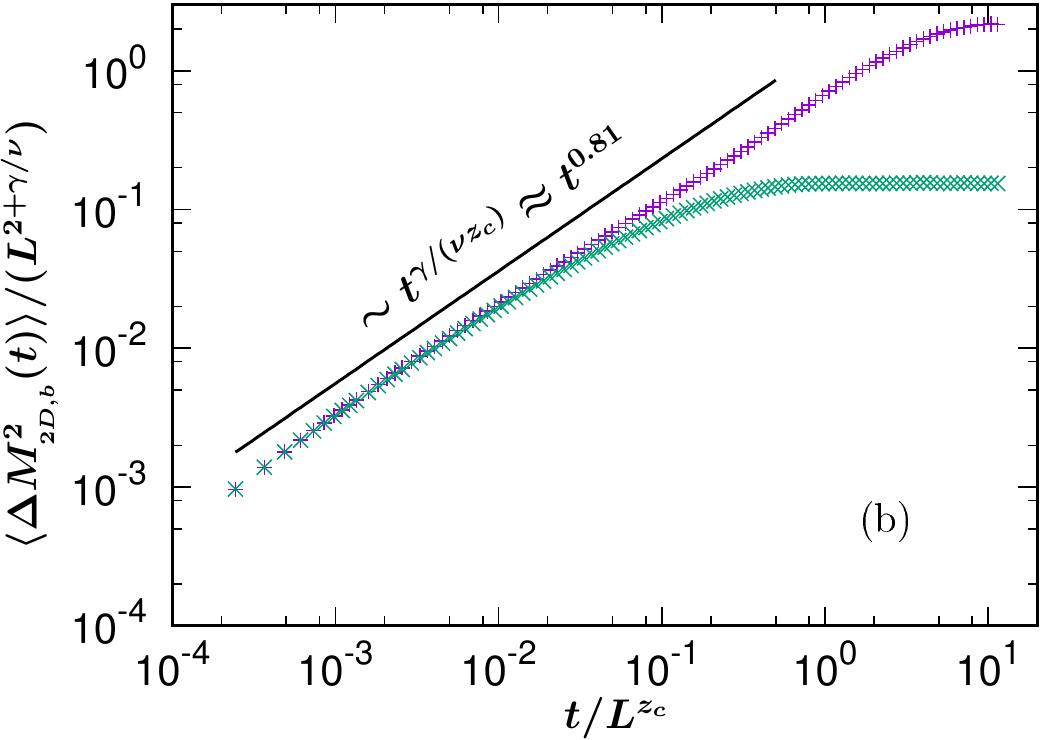}
  \caption{Comparison of the mean-square displacement for the line
    [Fig. (a), system size $L=128$], and bulk [Fig. (b), system size
      $L=64$] magnetisations for the Ising model in 2D, with periodic
    boundary conditions (magenta plusses) and free boundary conditions
    (green crosses). The data for the two different boundary
    conditions are on top of each other in the scaling regime,
    differing only at late times. \label{diffbound}}
\end{figure*}

In this appendix we demonstrate, in Fig. \ref{diffbound}, using two
examples that the deviations from the power-law behaviour at late
times, as seen in Fig. \ref{msd} are indeed caused by the periodic
boundaries.

\section*{Appendix B \label{sec7b}}

\setcounter{figure}{0}
\renewcommand{\thefigure}{B\arabic{figure}}

\begin{figure}[h] 
  \includegraphics[width=0.5\linewidth]{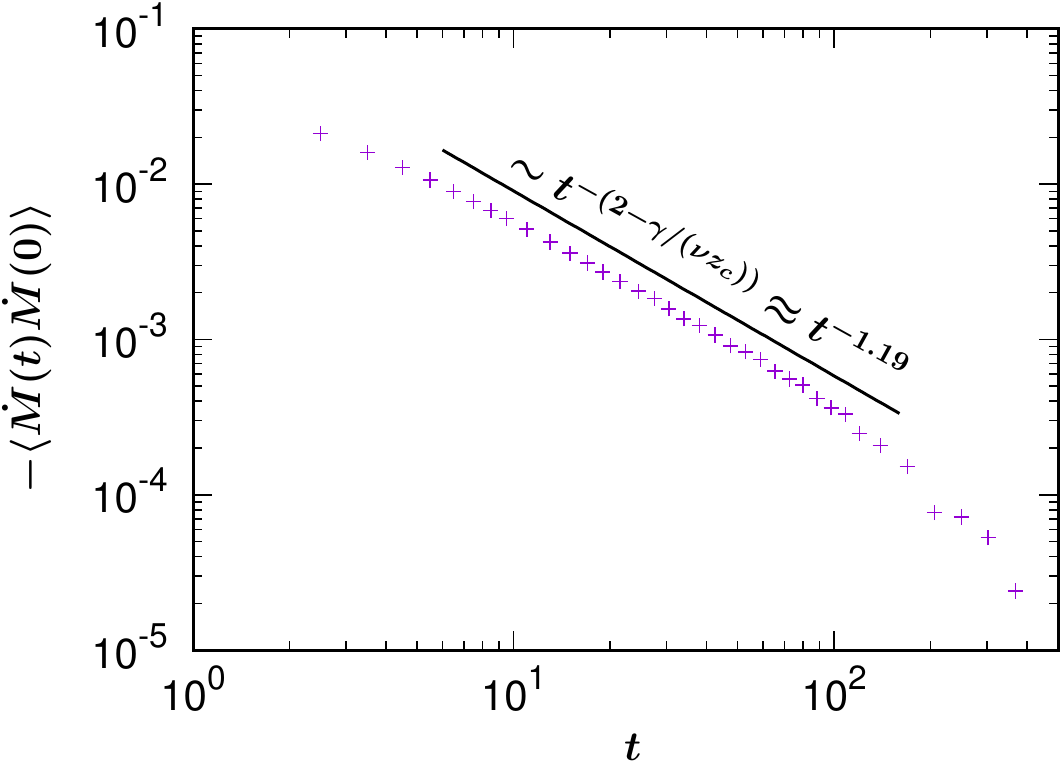}
   \caption{Velocity autocorrelation function $\langle
     \dot{M}(t)\dot{M}(0)\rangle$ of the 2D bulk magnetisation as a
     function of $t$. This quantity is negative, and behaves $\sim
     -t^{-2-\gamma/(\nu z)}\approx -t^{-1.19}$. The system size used
     in the simulation is $L=30$.}  \label{velauto}
\end{figure}

In this appendix, in Fig. \ref{velauto} we present a verification for
the Green-Kubo relation used to convert the vecolity autocorrelation
function (\ref{e4a6}) to anomalous diffusion (\ref{e4a7}): i.e., for
an anomalous diffusion exponent $c$ the velocity autocorrelation
function anomalous exponent must be $c-2$, as well as having an
overall negative sign in front.

\end{document}